\pdfoutput=1
\documentclass[10pt]{article}
\usepackage{graphicx}

\def\Title#1{\begin{center} {\Large #1 } \end{center}}
\def\Author#1{\begin{center}{ \sc #1} \end{center}}
\def\Address#1{\begin{center}{ \it #1} \end{center}}

\newcommand\pubblock{\rightline{\begin{tabular}{l} Proceedings of the Second Annual LHCP\\ \pubnumber\\
         \pubdate  \end{tabular}}}

\newenvironment{Abstract}{\begin{quotation} \begin{center} 
             \large ABSTRACT \end{center}\bigskip 
      \begin{center}\begin{large}}{\end{large}\end{center} \end{quotation}}

\newenvironment{Presented}{\begin{quotation} \begin{center} 
             PRESENTED AT\end{center}\bigskip 
      \begin{center}\begin{large}}{\end{large}\end{center} \end{quotation}}

\input{atlasphysics.sty}

\usepackage{color}
\usepackage[labelformat=simple]{subcaption}

\usepackage{siunitx}
\RequirePackage{lineno}

\newcommand\pubnumber{ ATL-PHYS-PROC-2014-109 }

\newcommand\pubdate{\today}

\def\affiliation{Center for High Energy Physics \\
University of Oregon, Eugene, OR 97403, U.S.A\\
\vspace{7mm} 
On behalf of the ATLAS Collaboration}

\begin{document}
\large
\begin{titlepage}
\pubblock

\vfill
\Title{Search for direct pair production of the top squark in all-hadronic final states in proton--proton collisions at $\rts=8\tev$ with the ATLAS detector}
\vfill

\Author{Chaowaroj Wanotayaroj}
\Address{\affiliation}
\vfill
\begin{Abstract}

The results of a search for direct pair production of the scalar partner to the top quark
using an integrated luminosity of \totalluminumnoerr~of proton--proton collision data at $\sqrt{s}=8$~TeV recorded with the ATLAS detector at the LHC are reported. The top squark is assumed to decay via $\stop \to t \ninoone$ or $\stop\to b\chinoonepm \to b  W^{\left(\ast\right)} \ninoone$, where $\ninoone$ ($\chinoonepm$) denotes the lightest neutralino (chargino) in supersymmetric models. The search targets a fully-hadronic final state in events with four or more reconstructed jets and large missing transverse momentum.  
No significant excess over the Standard Model background prediction is observed,
and exclusion limits are reported in terms of the top squark and
neutralino masses and as a function of the branching fraction of
$\stop \to t \ninoone$. For a branching fraction of $100\%$, top squark masses in the range
$270$--$645\GeV$ are excluded for $\ninoone$ masses below $30\GeV$.
For a branching fraction of $50\%$ to either $\stop \to t \ninoone$ or
$\stop\to b\chinoonepm$,  and assuming the $\chinoonepm$
mass to be twice the $\ninoone$ mass, top squark masses in the range
$250$--$550\GeV$ are excluded for $\ninoone$ masses below $60\GeV$.

\end{Abstract}
\vfill

\begin{Presented}
The Second Annual Conference\\
 on Large Hadron Collider Physics \\
Columbia University, New York, U.S.A \\ 
June 2-7, 2014
\end{Presented}
\vfill
\end{titlepage}
\def\thefootnote{\fnsymbol{footnote}}
\setcounter{footnote}{0}
%

\normalsize 


\section{Introduction}

The gauge hierarchy problem is a long standing puzzle for particle physics. The recent discovery of the Standard Model (SM)-like Higgs boson only re-emphasized the need to address this problem. One of the promising theories that can explain this issue is Supersymmetry (SUSY). The theory introduces supersymmetric partners of SM fermions and bosons. If the partner of the top quark (stop, \stop) has a mass below $\sim$1\nolinebreak\ TeV, then the divergence in the theoretical mass of the discovered Higgs boson due to the loop diagrams involving top quark can be mostly cancelled. Also, the theory provides a good candidate for dark matter.

\begin{figure}
	\centering
	\begin{subfigure}[b]{0.48\textwidth}
		\centering
		\includegraphics[width=0.5\textwidth]{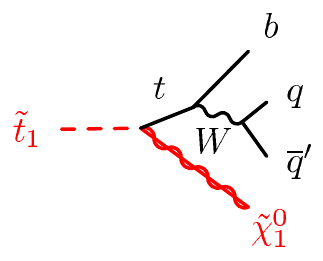}
		\caption{$\tone \rightarrow t \ninoone$.}
		\label{fig:stbqqN1}
	\end{subfigure}%
	~ 
	\begin{subfigure}[b]{0.48\textwidth}
		\centering
		\includegraphics[width=0.5\textwidth]{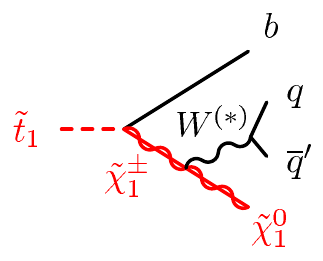}
		\caption{$\tone \rightarrow b \chinoonepm\rightarrow b  W^{\left(\ast\right)} \ninoone$.}
		\label{fig:stbqqN1-bC1.eps}
	\end{subfigure}
	\caption{Feynman diagrams illustrating the stop lighter mass eigenstate $\tone$ decay modes considered.}\label{fig:feynman}
\end{figure}

Given the mentioned scenario, one way to look for evidence for SUSY with the ATLAS detector~\cite{DetectorPaper:2008}\ is searching for direct stop pair production~\cite{OriginalPaper}. For this analysis, the top squark is assumed to decay via $\stop \to t \ninoone$ or $\stop\to b\chinoonepm \to b  W^{\left(\ast\right)} \ninoone$, where $\ninoone$ ($\chinoonepm$) denotes the lightest neutralino (chargino) in supersymmetric models, as shown in Fig.\ref{fig:feynman}.  The lightest supersymmetric partners (LSPs) will leave the detector undetected thus appeared as missing transverse momentum (\met). The advantage of the all-hadronic signature compared to those with leptons is the higher top branching ratio, less sensitivity to $\stop_L$ and $\stop_R$ mixing, and no neutrino contribution to the \met. On the other hand, reconstructing the tops are more ambiguous due to the number of jets. Also, jets are harder objects to measure compared to leptons in general.

\section{Signal Regions}

The basic signal selections are at least four distinct \antikt\ R=0.4 jets (two of which are b-tagged), no reconstructed electron or muon, and significant \met. To maximized the sensitivity to various models, the events are separated into different signal regions (SR). Signal Region A (SRA) is designed to target the events where all top daughters have been reconstructed as a jet thus requiring six or more jets, referred to as ``Fully Resolved'' events. SRA is further separated into SRA1-5 to target various $\tone$ and $\ninoone$ masses.

To enhance the sensitivity further, especially for the model with higher stop mass, the Signal Region B (SRB) includes the possibility of two or more top daughters merged into a single jet or mis-reconstructed in the detector, referred to as ``Partially Resolved'' events. SRB requires the event to contain four or five jets to ensure that there is no overlapped with SRA. SRB is further separated into two orthogonal signal regions using the top mass asymmetry variable \topmassasym, defined as:
\begin{equation}
\topmassasym = \frac{|\mantikttwelvezero - \mantikttwelveone|}{\mantikttwelvezero + \mantikttwelveone}.
\label{eqn:topmassasym}
\end{equation}
\mantikttwelvezero\ (\mantikttwelveone) is the mass of leading (sub-leading) reclustered jet (see Section~\ref{sec:topreco}). The lower value of \topmassasym\ indicates that the two top daughters are well separated thus the masses of both top candidates can be used in the selection. On the other hand, the higher value indicates that they are overlapped. Therefore, the SRB is divided into two regions: SRB1 where $\topmassasym < 0.5$ and the opposite is for SRB2.

Finally, Signal Region C (SRC) is designed to target the stop decay to a bottom quark and \chinoonepm. One of the quarks from the W decays is expected to be too soft to be detected as a jet. This requires exactly five jets in the events. SRC is also divided further into SRC1-3 to target different stop and \ninoone\ masses.

\subsection{Top Quark Reconstruction}
\label{sec:topreco}

Top quark decays are expect to be present in SRA and SRB. Since the event topologies differ, SRA and SRB employ different top reconstructions to maximize signal efficiency. For SRA, since the top daughters are fully reconstructed as jets, a top candidate is built by taking the two most likely b-jets as seeds and combine them with their closest two jets. This algorithm, referred to as $\Delta R_{min}$ method, has a very good signal acceptance with a reasonable background rejection.

Due to the merged or missing jets, a different algorithm is needed for SRB. A simple but effective method is to use \antikt\ with a bigger R parameter after pileup subtraction, and fully calibrated \antikt\ R=0.4 jets as inputs. Various R parameters have been studied and R=1.2 is chosen as the optimal choice for top tagging efficiency while retaining a reasonable backgrounds rejection.

\section{Backgrounds, Rejection and Estimations}

\begin{figure}
	\centering
	\begin{subfigure}[t]{0.48\textwidth}
		\centering
		\includegraphics[width=0.8\textwidth]{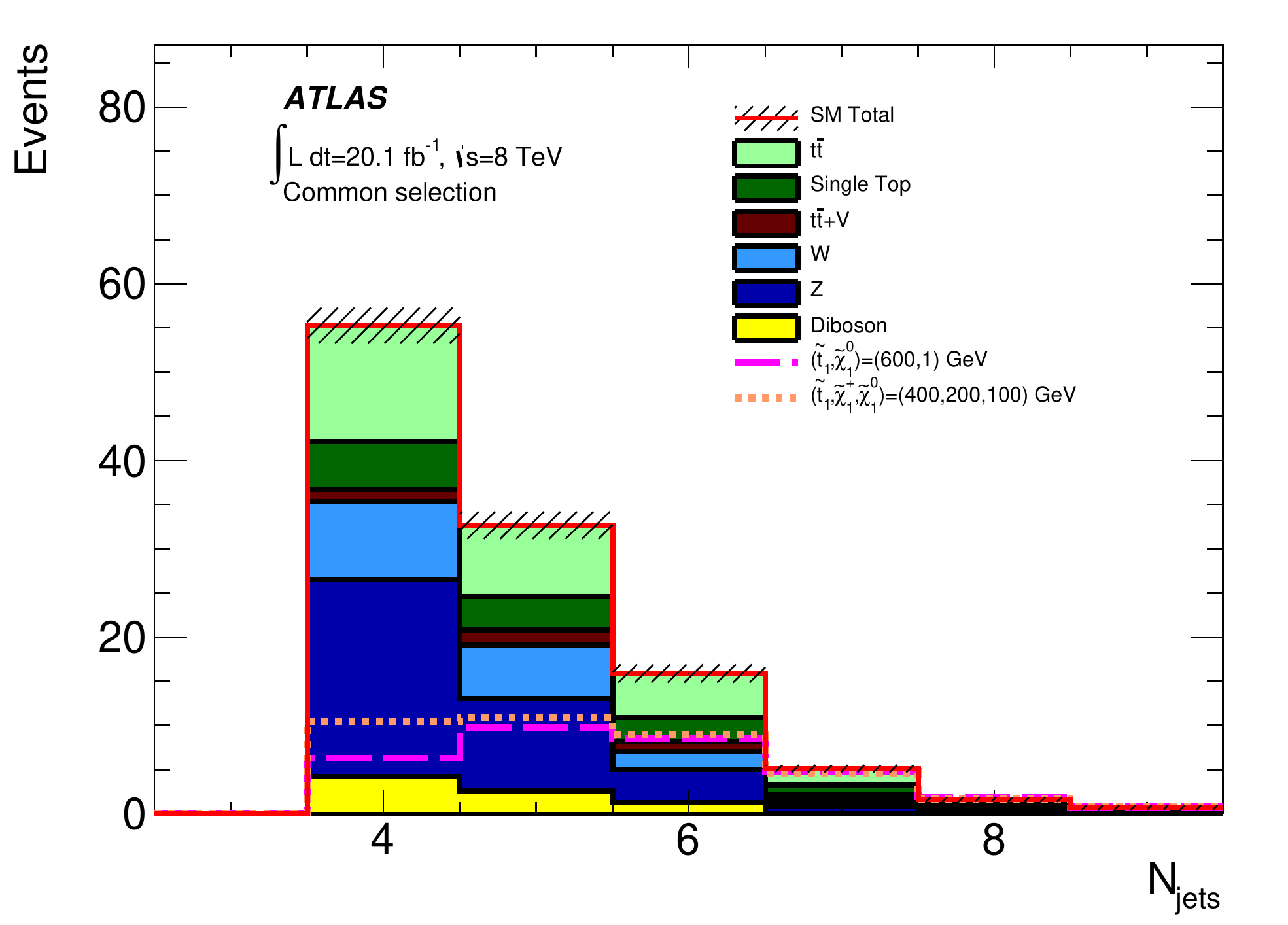}
		\caption{Distributions of the number
			of $R=0.4$ jets ($\pt > 35\GeV, |\eta| < 2.8$) for the dominant
			backgrounds and one signal case with $\met > 300\gev$ with common \mtbmetmindphi\ requirement. Data are not shown.
			}
		\label{fig:NJets}
	\end{subfigure}%
	~~~~~~~
	\begin{subfigure}[t]{0.48\textwidth}
		\centering
		\includegraphics[width=0.7\textwidth]{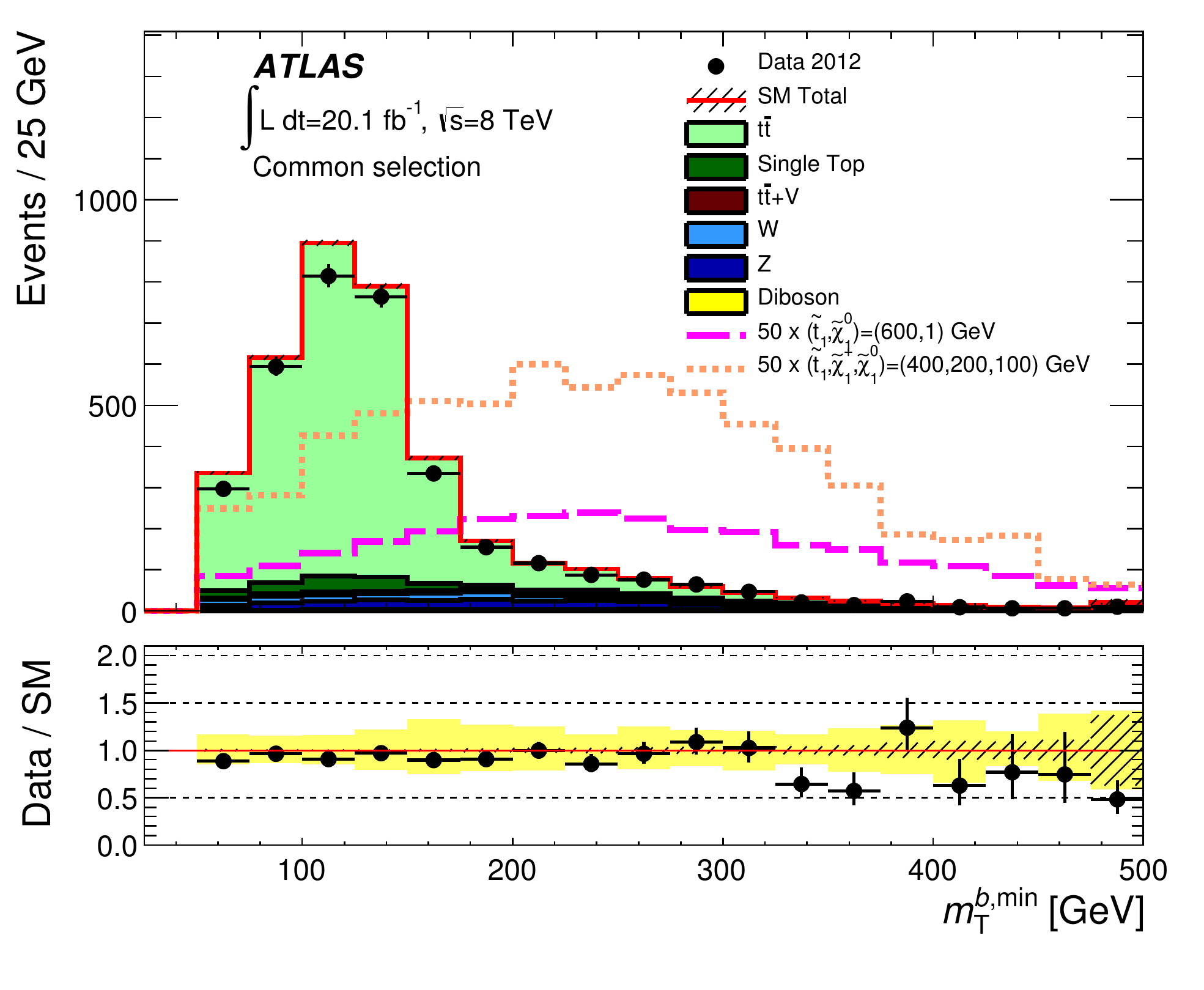}
		\caption{The distribution of \mtbmetmindphi\ in events with at least four jets excluding the requirement on \mtbmetmindphi. The ``Data/SM'' plot shows the ratio of data events to the total Standard Model expectation.}
		\label{fig:MtMetJetFinal}
	\end{subfigure}
	\caption{Distributions after the common selections. Backgrounds are shown as stacked, filled histograms.}\label{fig:mainplots}
\end{figure}

As shown in Fig.\ref{fig:NJets}, one of the major backgrounds is \ttbar. This occurs when one of the tops decays leptonically and jets are produced by initial or final state radiation. When plotting the transverse mass between the \met\ and closest b-jet, this background has a very sharp kinematic endpoint at around the top mass as shown in Fig.\ref{fig:MtMetJetFinal}. Therefore, the \mtbmetmindphi\ is required to be at least 175 \gev\ in all SRs.

The general strategy for background estimation is normalizing Monte Carlo (MC) to data in control regions (CR) for the three major backgrounds (\ttbar, W and Z with jets). The \ttbar\ background is estimated by using a \ttbar\ control region, which requires one reconstructed lepton treated as \met. Due to the requirement on the number of~\antikt\ R=0.4 jets, Z+jets with the Z boson decaying to neutrinos is the dominant background for SRB. The control region is defined using Z to two leptons which are treated as \met. W+jets, which is also a major background for SRB, has a control region similar to \ttbar\ but optimized for heavy flavor jets. Finally, the minor backgrounds (single top, tt+W/Z, Diboson, etc.) are estimated using MC only.

\section{Result and Interpretations}

The yields are showed in Table \ref{tb:SRresults}. There is no significant excess over SM background expectation observed. Therefore, the limit is set at the 95\% confidence level (CL) for the number of beyond-standard-model events.

\begin{figure}
	\centering
	\begin{subfigure}[t]{0.48\textwidth}
		\centering
		\includegraphics[width=0.65\textwidth]{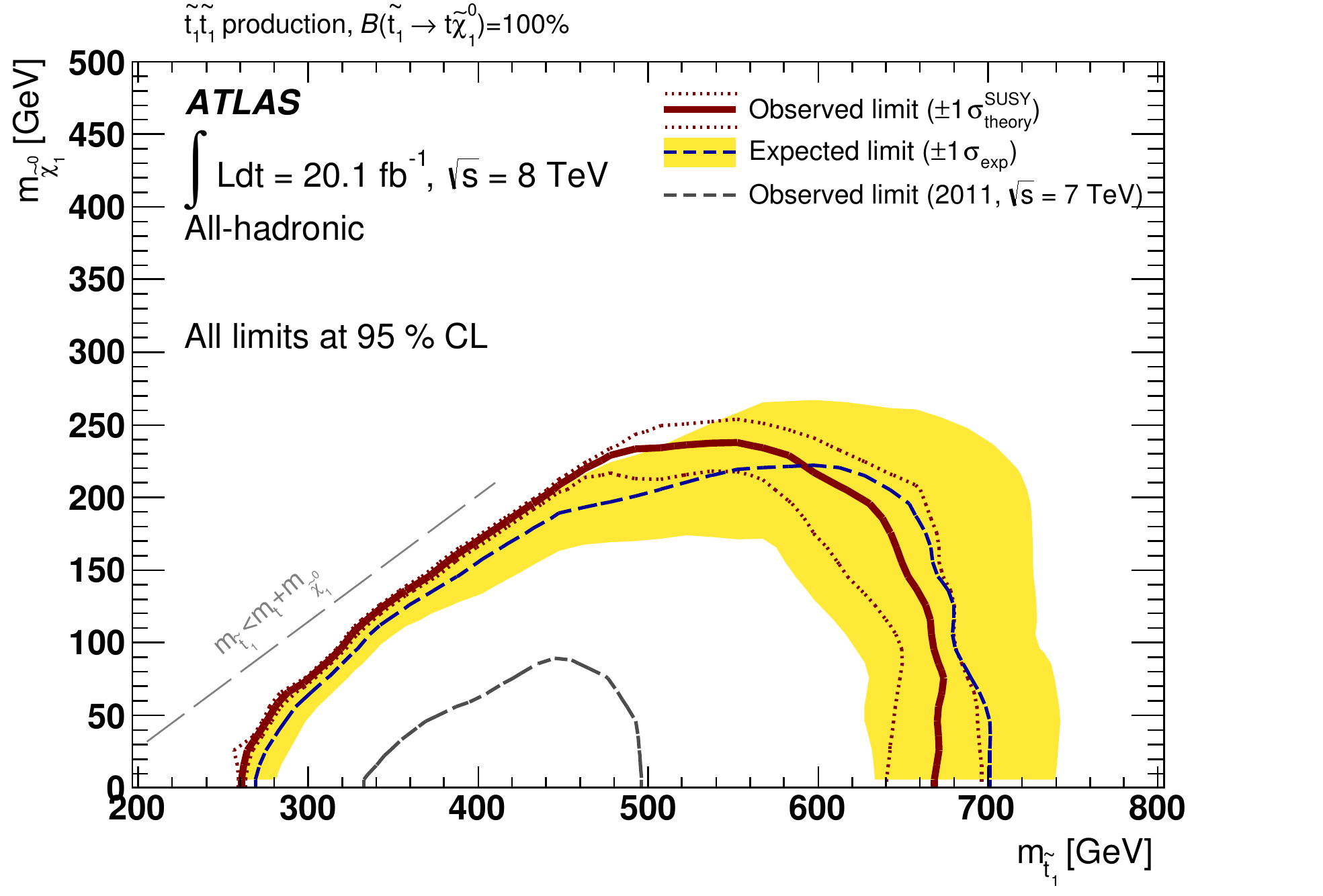}
		\caption{The scenario where both top squarks decay exclusively via $\stop \to t \ninoone$ and the top quark decays hadronically.}
		\label{fig:SRABCexclusion}
	\end{subfigure}%
	~~~~~~~ 
	\begin{subfigure}[t]{0.48\textwidth}
		\centering
		\includegraphics[width=0.65\textwidth]{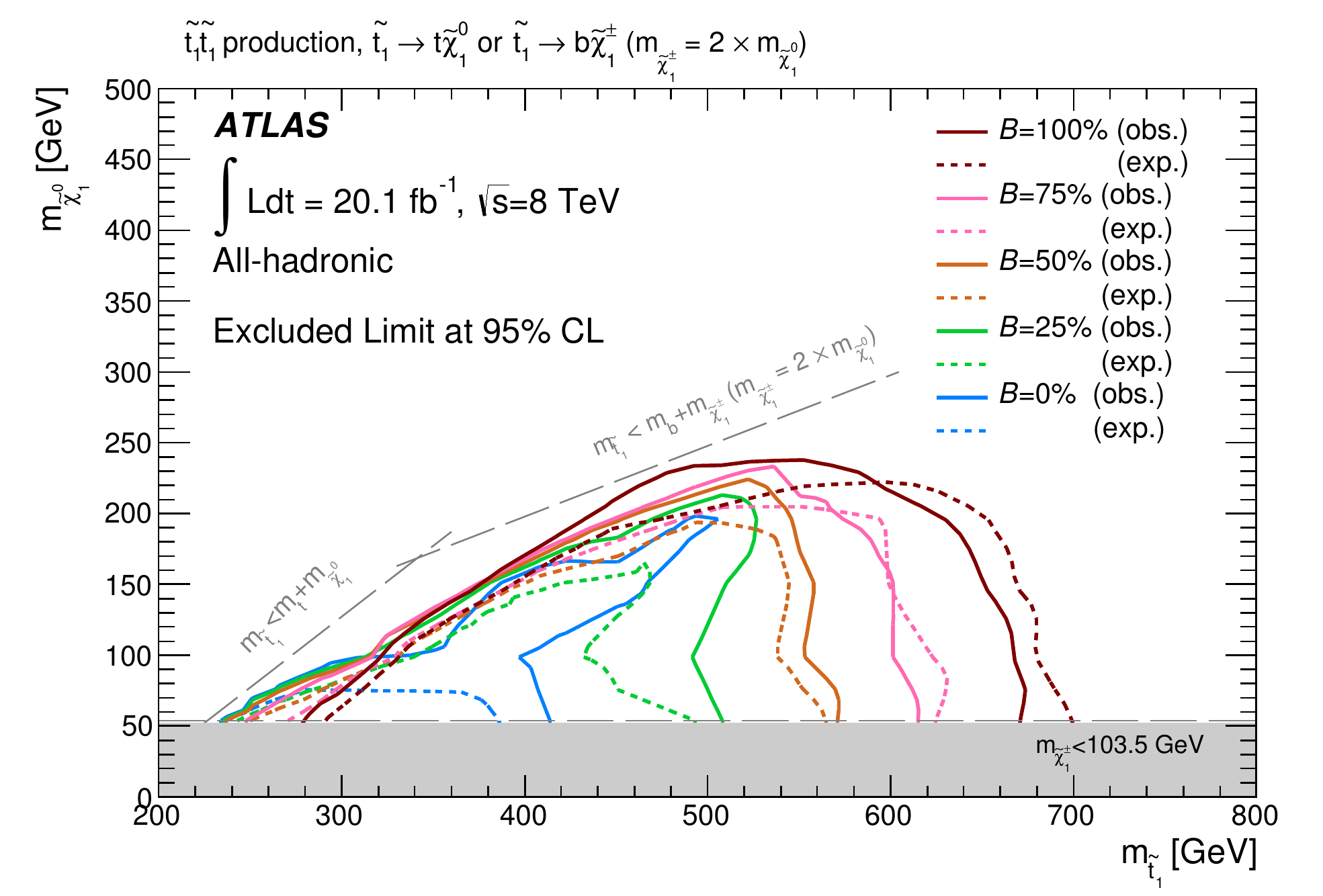}
		\caption{The scenario where the top squarks are allowed to decay via $\stop \to b \chinoonepm, \chinoonepm \to\Wboson^{\left(*\right)}\ninoone$. The $\chinoonepm$ mass is fixed to twice the $\ninoone$ mass. Each line colors are for different branching ratios of $\stop \to t \ninoone$}
		\label{fig:SRABCexclusion_BF}
	\end{subfigure}
	\caption{Exclusion contours at 95 $\%$ CL.}\label{fig:limitplots}
\end{figure}

Since SRA is orthogonal to both SRB and SRC, but SRB and SRC are overlapping, each of the SRA1-5 is first combined with SRB forming SRA+SRB combinations, and with each of SRC forming SRA+SRC combinations. Then for each $\tone$ and $\ninoone$ mass, the SRA+SRB or SRA+SRC with the best CL value is picked and used to plot exclusion contours. Fig. \ref{fig:SRABCexclusion} shows the scenario where only the $\stop \to t \ninoone$ is allowed while Fig.\ref{fig:SRABCexclusion_BF} shows the result when $\stop \to b \chinoonepm, \chinoonepm \to\Wboson^{\left(*\right)}\ninoone$ is allowed as well. The various color lines represent different branching ratios for $\stop \to t \ninoone$.

\begin{table}[h]
	\caption{Event yields in some signal regions (SRA, SRB, and SRC) are compared to the background estimate from the profile likelihood fit.}
	\begin{center}
\begin{tabular}{lcccccc} \hline\hline
 & \multicolumn{1}{c}{SRA1} & \multicolumn{1}{c}{SRA2} & \multicolumn{1}{c}{SRA3} & \multicolumn{1}{c}{SRA4}    & \multicolumn{1}{c}{SRB} & \multicolumn{1}{c}{SRC1}   \\ \hline
Observed events & $11$  & $4$& $5$& $4$                    & $2$     & $59$ \\ \hline \hline
Total SM & $\numRF{15.81}{3} \pm \numRF{1.90}{2}$ & $\numRF{4.05}{2} \pm \numRF{0.76}{1}$& $\numRF{4.07}{2} \pm \numRF{0.92}{1}$ & $\numRF{2.35}{2} \pm \numRF{0.66}{1}$ & $\numRF{2.38}{2} \pm \numRF{0.65}{1}$ & $\numRF{68.22}{2} \pm \numRF{7.34}{1}$\\ \hline
        $\ttbar$ & $\numRF{10.64}{3} \pm \numRF{1.90}{2}$ & $\numRF{1.81}{2} \pm \numRF{0.48}{1}$& $\numRF{1.05}{2} \pm \numRF{0.56}{1}$ & $\numRF{0.49}{2} \pm \numRF{0.34}{2}$   & \multicolumn{1}{l}{$0.10\;_{-\;0.10}^{+\;0.14}$}   & $\numRF{32.20}{2} \pm \numRF{4.42}{1}$         \\
        $\ttbar+\Wboson/\Zboson$ & $\numRF{1.80}{2} \pm \numRF{0.59}{1}$ & $\numRF{0.85}{2} \pm \numRF{0.29}{2}$ & $\numRF{0.82}{2} \pm \numRF{0.29}{2}$ & $\numRF{0.50}{2} \pm \numRF{0.17}{2}$     & $\numRF{0.47}{2} \pm \numRF{0.17}{2}$    & $\numRF{3.24}{2} \pm \numRF{0.82}{1}$        \\
        $\Zjets$ & $\numRF{1.42}{2} \pm \numRF{0.53}{1}$ & $\numRF{0.63}{2} \pm \numRF{0.22}{2}$& $\numRF{1.24}{2} \pm \numRF{0.39}{1}$ & $\numRF{0.68}{2} \pm \numRF{0.27}{2}$    & $\numRF{1.23}{3} \pm \numRF{0.31}{2}$   & $\numRF{15.70}{3} \pm \numRF{3.45}{2}$        \\
        $\Wjets$ & $\numRF{0.95}{1} \pm \numRF{0.45}{1}$ & $\numRF{0.46}{2} \pm \numRF{0.21}{2}$ & $\numRF{0.21}{2} \pm \numRF{0.19}{2}$ & \multicolumn{1}{l}{$0.06\;_{-\;0.06}^{+\;0.10}$}   & $\numRF{0.49}{2} \pm \numRF{0.33}{2}$    & $\numRF{8.48}{1} \pm \numRF{3.70}{1}$       \\
        Single top & $\numRF{1.00}{2} \pm \numRF{0.35}{1}$& $\numRF{0.30}{2} \pm \numRF{0.17}{2}$& $\numRF{0.44}{2} \pm \numRF{0.14}{2}$ & $\numRF{0.31}{2} \pm \numRF{0.16}{2}$    & $\numRF{0.08}{1} \pm \numRF{0.06}{1}$    & $\numRF{7.23}{2} \pm \numRF{2.92}{2}$        \\
        Diboson & \multicolumn{1}{c}{$<$ 0.4} & \multicolumn{1}{c}{$<$ 0.13}  & $\numRF{0.32}{2} \pm \numRF{0.17}{2}$ & $\numRF{0.32}{2} \pm \numRF{0.18}{2}$   & $\numRF{0.02}{1} \pm \numRF{0.01}{1}$    & $\numRF{1.14}{2} \pm \numRF{0.77}{1}$      \\
        Multijets & \multicolumn{1}{c}{$<0.001$} & \multicolumn{1}{c}{$<0.001$} &  \multicolumn{1}{c}{$<0.001$} & \multicolumn{1}{c}{$<0.001$}    & \multicolumn{1}{c}{$<0.001$}    & $\numRF{0.24}{2} \pm \numRF{0.24}{2}$     \\ \hline 
\end{tabular}
\label{tb:SRresults}

	\end{center}
\end{table}

\section{Conclusion}
The results of a search for direct top squark production with an all-hadronic experimental signature of jets and \met\ using an integrated luminosity of \totalluminumnoerr\ of proton--proton collision data at $\rts=8\TeV$ collected by the ATLAS detector at the LHC show no excess over SM expectations. Therefore, exclusion limit is set as a function of $\tone$ mass, $\ninoone$ mass, and $\stop \to t \ninoone$ branching ratio.


\begin{thebibliography}{99}


\bibitem{DetectorPaper:2008}
{\bf ATLAS} Collaboration, {\it {The ATLAS Experiment at the CERN Large Hadron
  Collider}},  {\em JINST} {\bf 3} (2008) S08003.

\bibitem{OriginalPaper} 
{\bf ATLAS} Collaboration,
{\it {Search for direct pair production of the top squark in all-hadronic final states in proton-proton collisions at $\sqrt{s}=8$ TeV with the ATLAS detector}}
 [\href{http://xxx.lanl.gov/abs/1406.1122}{{\tt arXiv:1406.1122}}].

\end{thebibliography}
\end{document}